\documentclass[twocolumn]{aastex631}

\usepackage{ragged2e}

\usepackage{color}

\def\m{\textit{$M_{\star}$}}
\def\msun{\textit{$M_{\odot}$}}
\def\cm{\textit{$\rm cm^{-2}$}}

\def\nz{\textit{$N_z$}}

\def\fp{\textit{$f_{\rm pair}$}}
\def\fd{\textit{$f_{\rm dual}$}}

\def\um{\textit{$\rm \mu m$}}
\def\zphot{\textit{$z_{\rm phot}$}}
\def\zspec{\textit{$z_{\rm spec}$}}
\def\logm{\textit{${\rm log}\,M_\star$}}

\def\lx{\textit{$L_{\rm X}$}}
\def\nh{\textit{$N_{\rm H}$}}
\def\ergs{\textit{$\rm erg\ s^{-1}$}}
\def\loglbol{\textit{${\rm log}\,L_{\rm bol}$}}
\def\lbol{\textit{$L_{\rm bol}$}}

\def\sex{{\tt SExtractor}}

\def\rp{\textit{$r_p$}}

\def\re{\textit{$R_{\rm e}$}}
\def\ss{\textit{$\rm S\acute{e}rsic$}}
\def\fagn{\textit{$f_{\rm agn}$}}

\def\ergs{\textit{$\rm erg\ s^{-1}$}}

\def\galfitm{{\tt GALFITM}}

\begin{document}

\title{Active Galactic Nuclei and Host Galaxies in COSMOS-Web. II. First Look at the Kpc-scale Dual and Offset AGN Population}

\author[0000-0002-1605-915X]{Junyao Li}
\affiliation{Department of Astronomy, University of Illinois at Urbana-Champaign, Urbana, IL 61801, USA}
\correspondingauthor{Junyao Li}
\email{junyaoli@illinois.edu}

\author[0000-0001-5105-2837]{Ming-Yang Zhuang}
\affiliation{Department of Astronomy, University of Illinois at Urbana-Champaign, Urbana, IL 61801, USA}

\author[0000-0003-1659-7035]{Yue Shen}
\affiliation{Department of Astronomy, University of Illinois at Urbana-Champaign, Urbana, IL 61801, USA}
\affiliation{National Center for Supercomputing Applications, University of Illinois at Urbana-Champaign, Urbana, IL 61801, USA}

\author{Marta Volonteri}
\affiliation{Institut d'Astrophysique de Paris, Sorbonne Universit\'e, CNRS, UMR 7095, 98 bis bd Arago, 75014 Paris, France}

\author[0000-0001-6627-2533]{Nianyi Chen}
\affiliation{McWilliams Center for Cosmology, Department of Physics, Carnegie Mellon University, Pittsburgh, PA 15213, USA}

\author[0000-0002-6462-5734]{Tiziana Di Matteo}
\affiliation{McWilliams Center for Cosmology, Department of Physics, Carnegie Mellon University, Pittsburgh, PA 15213, USA}
\affiliation{NSF AI Planning Institute for Physics of the Future, Carnegie Mellon University, Pittsburgh, PA 15213, USA}

\begin{abstract}
\noindent Kpc-scale dual and offset Active Galactic Nuclei (AGNs) are signposts of accreting supermassive black holes (SMBHs) triggered during late-stage galaxy mergers, offering crucial insights into the coevolution of SMBHs and galaxies. However, robustly confirmed systems at high redshift (e.g., $z>1$) are scarce and biased towards the most luminous and unobscured systems. In this study, we systematically search for kpc-scale (projected separation $<15$ kpc) dual and offset AGNs around 571 moderate-luminosity, X-ray-selected AGNs including the obscured population, utilizing deep HST ACS/F814W and multiband JWST NIRCam imaging from the COSMOS-Web survey. We identify 59 dual and 30 offset AGN candidates in late stage major mergers based on spatially-resolved spectral energy distribution analyses. This translates to $\sim28$ and $\sim10$ bona-fide dual and offset AGNs using a probabilistic pair counting scheme to minimize chance superpositions. 
Notably, the fraction of dual and offset AGNs among moderate-luminosity {($\lbol\sim10^{43}-10^{46}\ \ergs$)}, obscured AGNs is nearly two orders of magnitude higher than that of the most luminous, unobscured quasar pairs. We find tentative evidence for an increasing pair fraction among AGNs with redshift (from a few percent at $z\sim 0.5$ to $\sim22.9_{-17.7}^{+27.5}\%$ at $z\sim4.5$) and a higher occurrence rate of dual over offset AGNs. There is no pileup of dual/offset AGNs below $\sim 2~{\rm kpc}$ separations. These results generally align with predictions from the \texttt{ASTRID} and \texttt{Horizon-AGN} cosmological simulations when matching sample selection criteria, implying a high probability of both BHs being active simultaneously in late-stage major mergers. 
\end{abstract}

\section{Introduction}
Galaxy mergers play a pivotal role in shaping galaxy formation and evolution within the hierarchical structure formation paradigm. Since most massive galaxies harbor a central SMBH \citep[e.g.,][]{Kormendy2013}, the merge of two galaxies, facilitated by dynamical friction, should produce dual (on kpc-scale) and binary (in bound orbits) SMBHs \citep[e.g.,][]{Begelman1980, Yu2002, Chen2020dyn}. Mergers are proposed to significantly contribute to  SMBH growth and establish close connections with their host galaxies \citep[e.g.,][]{DiMatteo2005, Peng2007, Hopkins2008, Jahnke2011, Kormendy2013, Tang2023merger}. 
The enhancement of BH spin during galaxy mergers may boost the launch of relativistic jets \citep[e.g.,][]{Blandford1977, Breiding2024}, which are crucial for regulating a galaxy's star formation history through radio-mode feedback \citep[e.g.,][]{Heckman2014}. Moreover, the coalescence of massive binary BHs is the primary source of low-frequency gravitational waves (GW), whose detection is eagerly anticipated in the coming years with the Pulsar Timing Array (PTA) and Laser Interferometer Space Antenna and \citep[e.g.,][]{Kelley2017, Agazie2023}. Therefore, characterizing SMBH pairs in mergers provides crucial insights into the cosmological coevolution of SMBHs and galaxies and helps forecast GW emissions from binary SMBH inspiral and coalescence \citep[e.g.,][]{Steinborn2016, RosasGuevara2019, DeRosa2019, Chen2020dyn, Volonteri2022, Chen2023, LiK2023}. 

Since gas-rich mergers are expected to trigger intense gas inflows toward galactic centers, galactic-scale SMBH pairs can be detected as dual AGNs when both BHs are accreting, or as offset AGNs\footnote{In the literature, offset AGNs sometimes refer to a single AGN deviating from the center of its own host galaxy (i.e., off-center AGNs), which are likely caused by recoiling BHs after BH mergers. In this study, offset AGNs are when only one of the two SMBHs in a merging galaxy pair is active.} when only one nucleus is active \citep[e.g.,][]{Liu2011, Koss2012, Comerford2015, Silverman2020, Stemo2021, Yue2021, Tang2021, Shen2021, ChenY2023, Barrows2023}.  
Investigating dual and offset AGNs with kpc-scale ($\lesssim10$ kpc) separations at cosmic noon ($z\sim1-3$) is particularly important. This epoch marks the peak of cosmic star formation and AGN activity, during which tidal perturbations in the nuclear region induced by gas-rich mergers become more frequent and significant in fueling SMBH growth \citep[e.g.,][]{Blecha2018, Duncan2019}. The observed frequency of dual and offset AGNs as functions of redshift, separation, and host properties can elucidate our understanding of the AGN duty cycle and the dynamical evolution of SMBH pairs in mergers \citep[e.g.,][]{Shen2023}. These AGN pairs also set the initial conditions for binary SMBH coalescence at $z < 1$, which dominates the low-frequency GW background \citep[e.g.,][]{Agazie2023}, providing essential constraints on the anticipated GW emissions from merging SMBH pairs \citep[e.g.,][]{Chen2020dyn, LiK2023}.

Despite the keen interest in kpc-scale dual AGNs, their robust confirmation in the distant universe has been limited to a few cases due to stringent resolution demand and the limited duty cycle of dual AGNs \citep[e.g.,][]{Silverman2020, Shen2021, Tang2021, ChenY2023, Perna2023}. Recent advancements in the Varstrometry \citep{Shen2019, Hwang2020, Shen2021, ChenY2022, ChenY2023} and Gaia Multi Peak \citep{Mannucci2022, Ciurlo2023} selection techniques, which leverage Gaia's superb astrometry precision and excellent point spread function (PSF) in combination with large quasar catalogs like SDSS, have enabled efficient discoveries of candidate kpc-scale (subarcsec) dual quasars beyond $z\gtrsim1$. However, these techniques face significant challenges in distinguishing between dual and lensed quasars \citep{Gross2023, Li2023, Ciurlo2023}. The limited sensitivity of Gaia and SDSS, along with limitations inherent in the selection methods, tends to bias the selection towards the most luminous (bolometric luminosity $\lbol\gtrsim10^{46}\,\ergs$) and unobscured (obscuring column density $\nh\lesssim10^{22}\,\ergs$) quasar pairs. 
In fact, nuclear obscuration by gas and dust is ubiquitous and can be extreme during merger-induced starbursts, where substantial BH growth takes place. Most dual AGNs are thus likely to be optically type 2 and X-ray obscured \citep[e.g.,][]{Ricci2017, Blecha2018, Pfeifle2019, Ricci2021}. Consequently, the kpc-scale dual AGN fraction derived from Gaia and SDSS-selected type 1 quasar pairs ($\lesssim0.1\%$) is an order of magnitude lower than simulation predictions at $z\sim2$ \citep{Shen2023}, making them unrepresentative of dual AGNs in mergers. 

The advent of the James Webb Space Telescope (JWST) has revolutionized the study of AGNs across cosmic history. Its unparalleled sensitivity, spatial resolution, and wavelength coverage have enabled spatially-resolved analyses of AGNs and their host galaxies out to $z>6$, allowing for unprecedented characterization of AGNs in merging systems and their immediate surroundings. For instance, leveraging multiband NIRCam imaging data, \cite{cid42} demonstrated that CID-42, previously considered one of the most promising candidates for a GW recoiling BH, is actually a pair of merging galaxies hosting a single AGN. Furthermore, using deep NIRSpec IFU data from the GA-NIFS survey, \cite{Perna2023} reported spectroscopic confirmation of  four dual AGNs at $z\sim3.5$ (with one additional candidate) with separations ranging from $3-28$ kpc out of 17 AGNs at $2<z<6$, indicating a dual AGN fraction of $\sim20-30\%$. Coupled with the prevalence of companions around other JWST-observed high-redshift AGNs \citep[e.g.,][]{Matthee2023, Perna2023env}, these findings suggest that AGN in pairs in high-redshift, low-to-moderate-luminosity systems may be much more common than their high-luminosity, unobscured counterparts. 

In this work, we use deep HST and JWST imaging in the COSMOS field, which offers a large survey area ideal for discovering rare systems, to systematically search for typical (i.e., lower luminosity and obscured) kpc-scale dual and offset AGNs and to assess their evolution across cosmic times. Importantly, the spatial resolution of JWST can probe pairs down to $\sim1$~kpc scales without loss of completeness due to blending issues. The superb depth and multiwavelength coverage of HST+JWST will greatly facilitate companion detection, photometric redshift estimation, and AGN identification. In contrast, AGN pairs selected solely based on  HST imaging \citep{Stemo2021} face significant contamination from foreground and background stars and galaxies (see Section \ref{subsec:pair}). 
This paper is organized as follows. In Section \ref{sec:data} we describe the data used in this study. The strategy for identifying dual and offset AGN candidates is introduced in Section \ref{sec:selection}. The results are presented in Section \ref{sec:result} and summarized in Section \ref{sec:conclusion}. Throughout this paper we adopt a flat $\Lambda$CDM cosmology with $\Omega_{\Lambda}=0.7$ and $H_{0}=70\,{\rm km\,s^{-1}Mpc^{-1}}$. Magnitudes are given in the AB system.

\section{Data and Sample}
\label{sec:data}

\subsection{Imaging data and PSF model}
COSMOS-Web \citep[PIs: Kartaltepe \& Casey, program ID=1727]{Casey2023} is a 255-hour treasury JWST imaging survey covering 0.54 deg$^2$ in four NIRCam filters (F115W, F150W, F277W, and F444W), achieving a $5\sigma$ point source depth of $27.5-28.2$ magnitudes. Additionally, MIRI imaging over a 0.19 deg$^2$ region is conducted in parallel in the F770W filter. In this study, we utilize the fully reduced NIRCam imaging data from \cite{Zhuang2024}, which encompasses all COSMOS-Web observations up to May 2023, covering a survey area of $\sim0.26$ deg$^2$. Additionally, we complement the NIRCam imaging data with HST/ACS imaging (0\farcs03 pixel$^{-1}$, same as NIRCam data) in the F814W band from the CANDELS survey using the publicly released v1.0 mosaics \citep{Koekemoer2011}.

The details on data reduction and PSF construction are presented in \cite{Zhuang2024}. Here we briefly summarize our method. We use the software \texttt{PSFEx} \citep{Bertin2011} to construct point spread function (PSF) models, utilizing all high signal-to-noise ratio ($>100$) point-like sources in the NIRCam mosaics. For HST data, PSF models are constructed using stars within a radius of 1\farcm5 of the AGN target (Section \ref{subsec:xray_sample}) in  $5^\prime \times 5^\prime$ cutouts. The error image generated from the \texttt{resample} step of the JWST pipeline is used as the sigma image in imaging analysis. Since only weight images (inverse variance) are available for ACS F814W mosaics, we construct sigma images by adding Poisson noise to the background noise (derived from the weight image), accounting for cosmic ray flagging, flat fielding, and correlated noise from resampling.

\begin{figure*}
\includegraphics[width=0.49\linewidth]{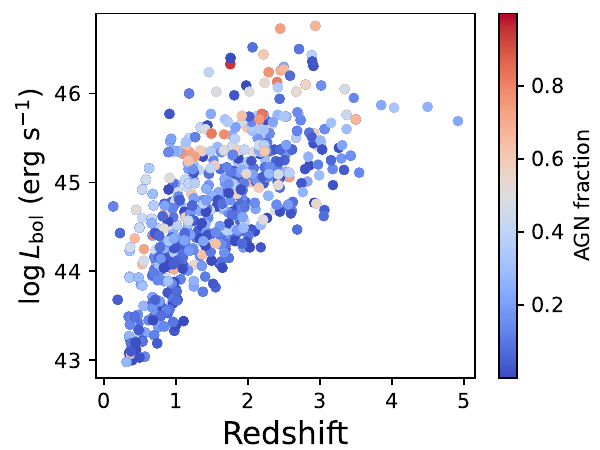}
\includegraphics[width=0.485\linewidth]{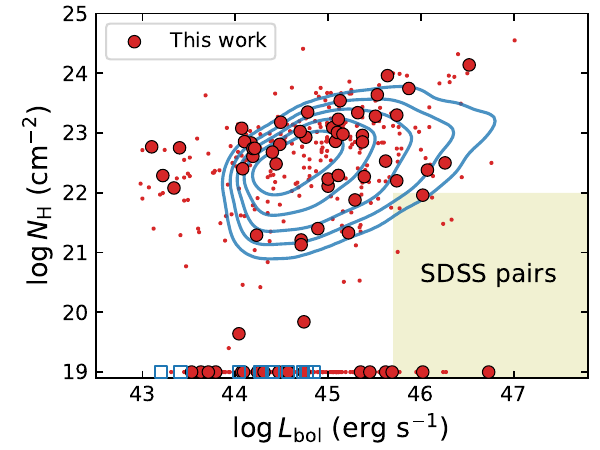}
\caption{Left: bolometric luminosity vs. redshift for the parent sample of 571 AGNs. Each data point is color-coded by the decomposed AGN to total flux ratio derived in the F277W band (Section \ref{subsec:galfitm}). Right: column density vs. bolometric luminosity for the 78 AGNs (red circles) in the final pair sample, plotted on top of all 571 AGNs (red dots). AGNs in pairs without \nh\ measurements are marked by blue squares. The blue contours display AGNs with $z>2$ in the \texttt{ASTRID} simulation that match the detection limit in \lbol\ of our sample. The parameter space covered by pairs selected based on SDSS type 1 quasars at $z\sim2$ is shown as a shaded region, assuming optically type 1 quasars correspond to X-ray $\nh < 10^{22}\ \cm$ \citep[e.g.,][]{Silverman2005}.}
\label{fig:lumin}
\end{figure*}

\subsection{X-ray AGN Sample}
\label{subsec:xray_sample}
Our strategy for searching for dual and offset AGNs is based on detecting potential companions around known AGNs. We adopt X-ray AGNs from the Chandra COSMOS-Legacy survey  \citep{Civano2016, Marchesi2016}, a deep 4.6 Ms Chandra program covering 2.15 deg$^2$ of the COSMOS field, with a flux limit of $\sim8.9\times 10^{-16}$ erg s$^{-1}\ \cm$ in the 0.5--10 keV band. Out of 4016 X-ray detected sources, 571 AGNs ($\lx \gtrsim 10^{42}\ \ergs$) and their surroundings fall within the footprint of public NIRCam imaging in \cite{Zhuang2024} in all four bands, constituting our parent AGN sample. This sample includes both unobscured and obscured AGNs (based on X-ray absorption column density, rather than optical spectroscopic classification). 

The identification of optical counterparts for the X-ray AGNs is provided in \cite{Marchesi2016}. However, when the AGN has a close companion, correctly assigning the X-ray source to the right optical counterpart becomes challenging due to the limited resolution and sensitivity of ground-based imaging and Chandra. {In case of ambiguity, we identify the correct JWST counterpart as the source showing the strongest AGN signatures in multiband images and nuclear SEDs (see Sections \ref{subsec:galfitm} and \ref{subsec:pair}). Twenty objects were found to have incorrect optical counterparts in \cite{Marchesi2016}}. 

Reliable spectroscopic redshifts (\zspec) are available for 314 objects (\texttt{q\_spec} $ = 2.0$) after updating the counterpart information. Photometric redshifts (\zphot) are also provided in \cite{Marchesi2016}, derived from multiwavelength data in the COSMOS field using \texttt{LePhare} \citep{Ilbert2009}, following the methodology outlined in \cite{Salvato2011}. However, for close pairs with separations less than $\sim1\arcsec$, the ground-based photometry used to estimate \zphot\ is susceptible to contamination from nearby companions, as most are not adequately deblended in the multiwavelength dataset used in \cite{Marchesi2016}. Therefore, we conduct our own \zphot\ measurements using deblended HST+JWST photometry for both AGNs and their companions (see Section \ref{subsec:photoz}). 

We compile the intrinsic, absorption-corrected X-ray luminosity (\lx) and column density (\nh) of obscuring materials (measured using the redshifts in \citealt{Marchesi2016}) for our targets from \cite{Lanzuisi2018}, \cite{Marchesi2016spec}, and \cite{Civano2016}, prioritizing the former for common sources. The bolometric luminosity (\lbol) is then derived from \lx\ using the bolometric conversion factor in \cite{Lusso2012}. {Among our sample, 218 are X-ray unobscured ($\nh<10^{22}\ \cm$), 239 are X-ray obscured ($\nh>10^{22}\ \cm$), and 114 do not have $\nh$ measurements due to insufficient counts. For the 20 X-ray AGNs with mismatched optical-near-infrared (NIR) counterparts, their \lx\ and \nh\ values, which are mainly used to illustrate the general sample properties, are taken as reported without refitting the X-ray spectrum using the updated redshift.}

The distribution of the 571 AGNs on the $\lbol-z$ and $\lbol-\nh$ planes is shown in Figure \ref{fig:decomp}. Our sample reaches \lbol\ values $\sim2$ orders of magnitude lower than the pair sample selected based on type 1 quasars \citep[e.g.,][]{Silverman2020, Shen2023}, and includes both unobscured and heavily obscured AGNs. The apparent positive correlation between \nh\ and \lx\ is partly due to a selection bias, as only luminous objects can be detected in cases of heavy obscuration. The broad coverage in luminosity and obscuration makes our sample ideal for exploring more typical dual and offset AGN systems, although it remains biased against Compton-thick AGNs, where even hard X-ray photons can be significantly absorbed and Compton scattered \citep[e.g.,][]{Lanzuisi2018, Li2019}.

\section{Methods}
\label{sec:selection}
Our selection of dual and offset AGN candidates involves a multi-step procedure. First, we run source detection using \sex\ \citep{Bertin1996} to identify sources within the $400\times400$ pixel ($12\arcsec \times12\arcsec$) cutouts, targeting potential companion galaxies around the central AGN. We then perform multiwavelength simulataneous AGN-host decomposition using \galfitm\ \citep{Haussler2013} to disentangle the blended flux of the central AGN, its host galaxy, and close companions. This enables us to measure their \zphot\ and stellar mass to assess the physical association of projected pairs and the merger mass ratio. Additionally, we evaluate whether the companion may contain a second AGN by analysing its spectral energy distribution (SED) in the nuclear region resolved by HST and JWST. 

We note that these steps are iterative, as we found that some X-ray AGNs were assigned to incorrect optical counterparts in \cite{Marchesi2016} by examining their images and nuclear SEDs. The results are eventually combined to select the final candidate dual and offset AGN sample with a projected separation $r_p<15$ kpc and a merger stellar mass ratio $>1:6$ in Section \ref{sec:result}. Note that CID-42 has been previously analyzed in \cite{cid42}, where it was confirmed to be a merging pair of galaxies hosting a single AGN (i.e., an offset AGN). We do not repeat the analysis here and directly include it in our offset AGN sample.

\subsection{Source detection}

We select F277W as the reference band for source detection using \sex\ due to its improved spatial resolution compared to F444W and its insensitivity to star- forming clumps and dust extinction effects compared to bluer bands. We adopt \texttt{deblend\_nthresh} = 3.0, \texttt{deblend\_mincont} = 0.005, \texttt{detect\_minarea} = 5 to identify companion galaxies located more than $0.\arcsec5$ away from the central AGN. The relatively high value of \texttt{deblend\_mincont} is selected to prevent the outer clumpy disk from being segmented into mutiple sources. Within the central $0.\arcsec5$ region,  we adjust \texttt{deblend\_mincont} to 0.001 to improve the deblending of close cores. Additionally, we conduct source detection in the F150W band with \texttt{deblend\_mincont} = 0.001 to deblend close pairs that may not be easily resolvable in F277W. \sex-detected sources are labelled for each AGN-companion system, but the primary AGN is not necessarily labelled as source ``0'' (see Table~\ref{table:decomp}.) 

We find that, in most cases, the source detection results from the three runs are consistent, although discrepancies occur in certain instances. These discrepancies can be attributed to several factors: 1) the central, closely-separated nuclei are not successfully deblended in F277W due to its lower resolution compared to bluer bands; 2) the star-forming clumps across galaxy disks and spiral arms, as commonly observed in our sample, are segmented into multiple sources in F150W; and 3) some faint companions are blended with the outer disk of the central AGN, and adopting either \texttt{deblend\_mincont} = 0.005 or 0.001 in F277W does not effectively deblend them. 
Unfortunately, there is no universal set of parameters capable of correctly deblending all cases. Therefore, we conduct a visual inspection of the inconsistent source detection results and manually identify the correct ones. 

A significant challenge in identifying genuine pairs in deep JWST images is the frequent presence of foreground and faint background sources near the X-ray AGN. To mitigate contamination from objects unrelated to the central AGN, we require the projected separation between the AGN and its companion to not exceed $\sim15$ kpc ($1.\arcsec8$ at $z\sim2$), where \rp\ is derived using either reliable spectroscopic redshifts or \zphot\ of the central AGN measured in Section \ref{subsec:photoz}.  
We also impose an initial cut on the \texttt{MAG\_AUTO} of the companion, requiring it is no more than 3 mags fainter than the decomposed host galaxy magnitude of the central AGN in the F277W band (see Section \ref{subsec:galfitm}). These criteria minimize the likelihood of chance superpositions and select systems likely in the late merging phase, during which the global galactic environment becomes highly disturbed and enhances the probability of triggering dual AGNs. A sample of 163 AGNs with their 197 relatively bright close companions survived these criteria and constitute our parent pair sample.

\subsection{Multiwavelength simultaneous AGN-host image decomposition}
\label{subsec:galfitm}

\begin{figure*}
\includegraphics[width=\linewidth]{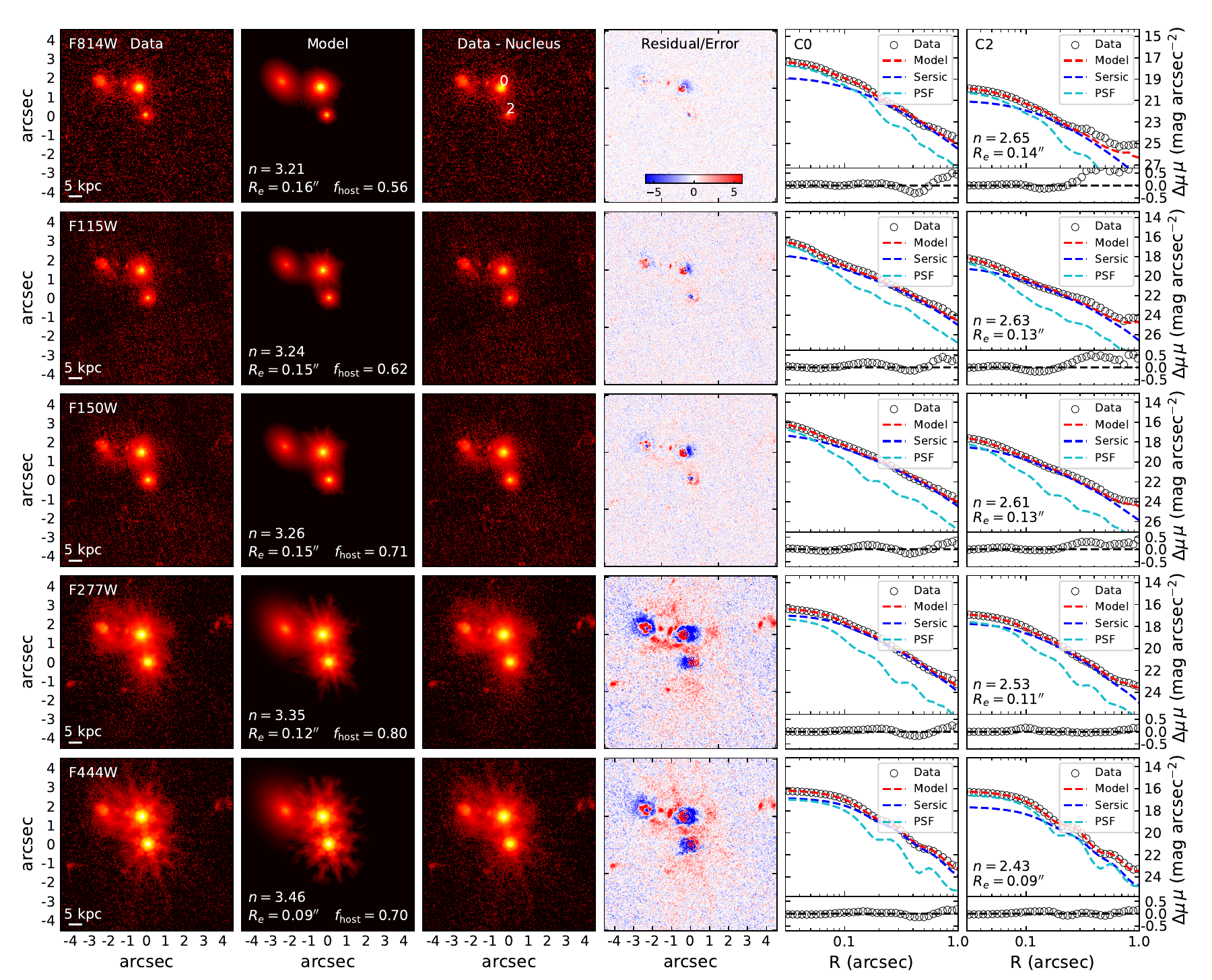}
\caption{Multiwavelength simultaneous AGN-host decomposition result for the dual AGN candidate CID-252 at $\zspec=1.13$. CID-252-0 (labeled in the third column in the first row) is the X-ray AGN, CID-252-2 is the companion associated with the system based on photo-$z$ probability metrics (Section \ref{subsec:photoz}), and the other companion is likely a background interloper with $\zphot \sim 2.9$. Both nuclei require a \ss\ + PSF model to properly fit their emissions. The columns are (1) observed data, (2) best-fit model, (3) PSF-subtracted data (i.e., galaxy emission only), (4) residual map, (5-6) radial surface brightness profile of CID-252-0 (C0) and CID-252-2 (C2). }
\label{fig:decomp}
\end{figure*}

We conduct multiwavelength simultaneous AGN-host image decomposition using \texttt{GALFITM} \citep{Haussler2013} to separate emissions from multiple components \citep[e.g.,][]{Zhuang&Ho2023, Zhuang2024}. \texttt{GALFITM} is a multiwavelength version of \texttt{GALFIT} \citep{Peng2010}, designed to account for wavelength-dependent galaxy structures. For each AGN and companion galaxy, we generate individual cutout images, positioning each source at the image center. Sources blended with the central target or other companions within the cutouts are fitted simultaneously, while unblended objects are masked. The cutout size is automatically determined to cover the segmentation map of all blended sources to be fitted within the cutout.

Given the wide range of redshifts and diverse morphologies of our objects, we use a consistent fitting approach that includes a PSF model to fit the AGN and a single \ss\ model for its host galaxy.  Although some objects display notable tidal features, our primary objective is to measure the flux of the AGN and its host galaxy within a common area for SED analysis. Missing the tidal features across all bands does not significantly impact the measurement of integrated flux. By default, the companion is fit with a \ss\ model. Additionally, we augment each \ss\ model with a PSF model for companions that likely contain a second AGN identified in Section \ref{subsec:second}. 

The centers of the AGN and the host galaxy are tied together and allowed to vary independently across different bands to account for residual astrometry mismatches ($\lesssim 0.3$ pixel among NIRCam images and $\sim1.3$ pixel between HST F814W and NIRCam F277W). We adopt constant ellipticity and position angle across wavelengths, while allowing the magnitude of the \ss\ model and PSF model to vary freely. The \ss\ index $n$ and half-light radius \re\ are permitted to vary linearly with wavelength. The \ss~index is restricted between 0.3 to 7, and the half-light radius is allowed to vary between 0.5 and 100 pixels, corresponding to $\sim0.07-25$~kpc across our sample's redshift range. The decomposition results, along with the nuclear SED fitting results in Section \ref{subsec:pair}, allow us to identify the correct optical-NIR counterpart of the X-ray AGN.

In Figure~\ref{fig:decomp}, we present the multiwavelength simultaneous AGN-host image decomposition results for the dual AGN candidate CID-252 (Section \ref{subsec:second}). This system requires an additional PSF model to fit the image of the companion, as otherwise, significant residuals remain in the central region, causing unphysically small \re. CID-252-0 is identified as the primary AGN due to its higher PSF luminosity. In most cases, the surface brightness profile of the companion is dominated by the \ss\ component, even for companions identified as dual AGN candidates in Section \ref{subsec:second} where we added an additional PSF model.  However, this does not necessarily imply that they do not harbor an AGN. In Figure~\ref{fig:lumin}, we color-code each of the 571 X-ray AGNs in our parent sample according to its decomposed AGN-to-total flux ratio ($f_{\rm agn}$) in F277W. It is evident that, unlike luminous broad-line quasars where the central bright point source typically dominates the emission, X-ray AGNs detected in deep Chandra exposures are mostly host-dominated with $\fagn \lesssim 20\%$, especially for obscured systems. The faint central AGNs in the rest-frame optical bands, either due to intrinsically lower luminosities or significant dust extinction, pose a challenge in identifying AGN activity based solely on image decomposition.  

\begin{figure*}
\includegraphics[width=\linewidth]{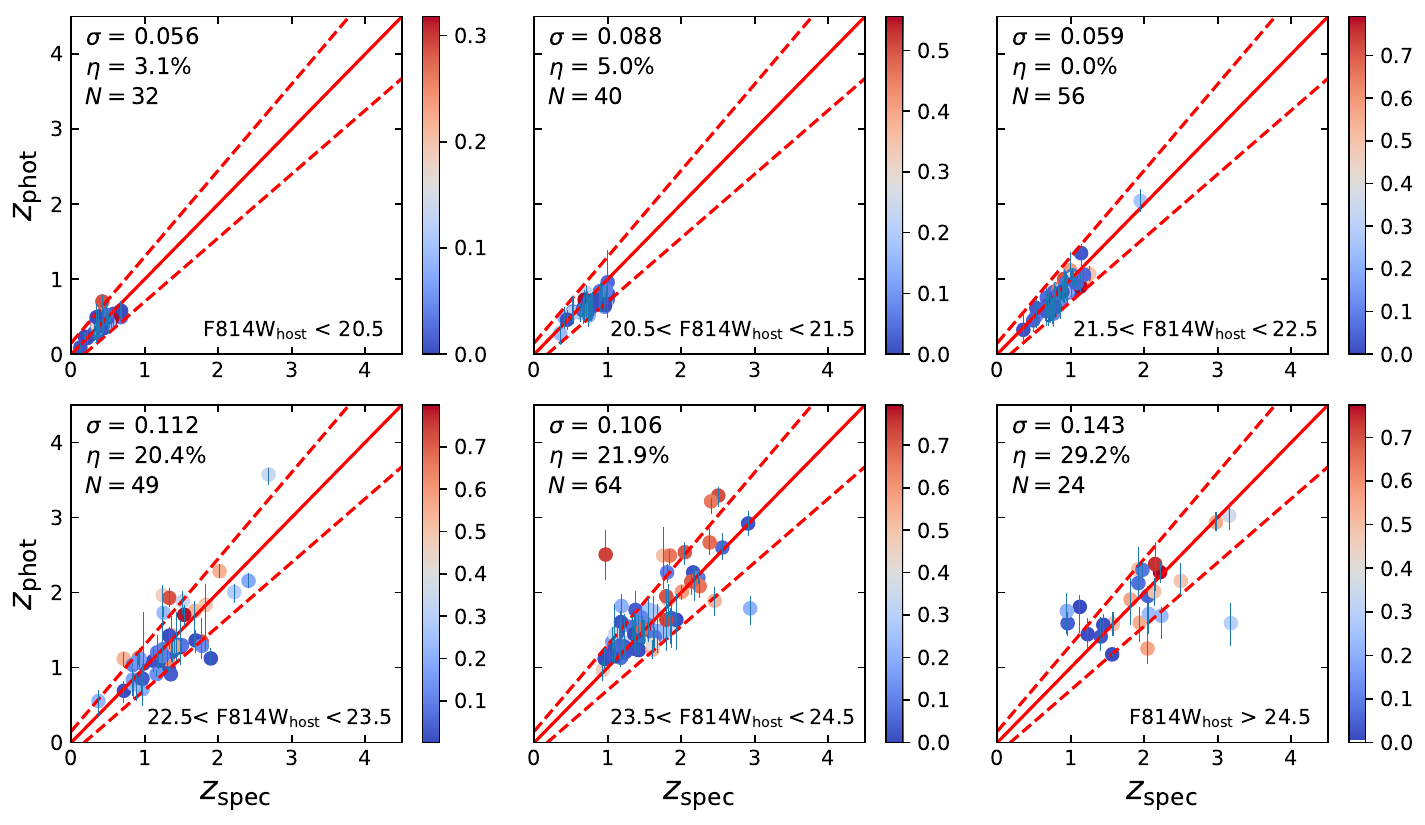}
\caption{Comparison between photometric redshifts and secure spectroscopic redshifts for AGNs with $f_{\rm gal}>20\%$, divided in bins of the F814W magnitude of their host galaxy. Each data point is color-coded by the decomposed AGN to total flux ratio. Red solid lines represent $\zphot=\zspec$, while red dashed lines represent $\zphot = \zspec \pm 0.15\times(1+\zspec)$. The accuracy ($\sigma$) and outlier fraction ($\eta$) of \zphot\ estimates, along with the number of objects in each magnitude bin ($N$) are labeled in each bin.}
\label{fig:lephare}
\end{figure*}

\subsection{Identification of dual AGN candidates}
\label{subsec:second}
To determine if the companion galaxy may host a secondary AGN (indicating a dual AGN system), we measure the multiband aperture flux within a 3-pixel ($0.\arcsec09$) radius around each companion to construct the SED of their nuclear region, minimizing host galaxy contamination \citep[e.g.,][]{cid42}. The PSF and \ss~components not centered on the companion of interest are subtracted when measuring the nuclear flux. The resolution of each image is matched to the F444W band, which has the lowest resolution, using a matching kernel derived from Fourier transform \citep{Aniano2011} available in {\tt{photutils}} \citep{Bradley2022}. The {\tt{CosineBellWindow}} is adopted to filter the high-frequency noise, with the $\alpha$ parameter set to 0.3, 0.45, and 0.9 for F814W, F115W--F277W, and F444W, respectively \citep{Rieke2023}. 

We then proceed with SED fitting using CIGALE v2022.1 \citep{Boquien2019, Yang2020} to identify the potential presence of an AGN in the companion galaxy. Initially, we fit the nuclear SED of each companion with pure stellar templates, fixing the redshift to that of the central AGN -- using either reliable \zspec\ in \cite{Marchesi2016} or \zphot\ derived in Section \ref{subsec:photoz}. This approach is based on the premise that if the system is indeed a dual AGN,  the companion should be at the same redshift. The \cite{BC03} stellar population model, \cite{Chabrier2003} initial mass function, and \cite{Calzetti2000} extinction law with $E(B-V)=0.0 - 1.0$ mag are assumed. We adopt a delayed star formation history with a stellar age ranging from 0.5 -- 9.0 Gyr and an e-folding time of the main stellar population varying from 0.1 -- 9.0 Gyr. 

We then incorporate the \texttt{SKIRTOR} AGN model \citep{Stalevski2016} into the fitting, fixing all parameters to default values except the inclination angle (30$^\circ$ for type 1 and $70^\circ$ for type 2) and AGN fraction (0.0--0.99).  The significance of the fitting improvement with an additional AGN component is quantified using the Bayesian Information Criteria (BIC). The BIC value is defined as ${\rm BIC} = k\, {\rm ln}(N) - 2 {\rm ln}(L)$, where $L$ is the maximized value of the likelihood function, $N$ is the number of data points, and $k$ is the number of free parameters. The BIC approach penalizes models with more free parameters when assessing the fitting improvement. 
Assuming Gaussian uncertainties, the likelihood is related to $\chi^2$ as $L = \rm exp\,(-\chi^2/2)$. Therefore, ${\rm \Delta BIC}$ can be expressed as $\Delta k {\rm ln}(N) + \Delta \chi^2$, where $\rm \Delta BIC$ and $\Delta \chi^2$ denote the differences in BIC and $\chi^2$ values calculated without and with the inclusion of additional AGN templates, respectively. A $\Delta \rm BIC>10$ is typically regarded as a significant improvement in the fitting result \citep{Raftery1995}. 

Given the redshift range of our sample and their obscured nature, identifying AGN activity relies primarily on detecting the excess of hot dust emissions from the AGN torus in the longest wavelength F444W band \citep[e.g.,][]{cid42}. While galactic-scale dust emission peaks at much longer far-infrared wavelengths and is overall negligible at rest-frame $\lambda \lesssim 3\ \um$ \citep[e.g.,][]{Yamada2023}, we consider its possible contribution to this excess emission, especially for low-redshift ($z\lesssim0.4$) systems where the 3.3~\um\ PAH emission coincides with the F444W band. To address this, we incorporate the \cite{Dale2014} dust emission model into our analysis. This model involves only one free parameter, with $\alpha$ set to [1.5, 2.0, 2.5] and the AGN fraction fixed to zero. 

We opted not to include the MIRI data in the SED fitting, as it overlaps with only a small portion of the current COSMOS-Web survey and is much shallower than the NIRCam imaging. Addintionally, there is a trade-off between wavelength coverage and spatial resolution. While including the F770W data might improve AGN selection through hot dust emissions, its spatial resolution is $\sim2-6$ times worse than NIRCam, leading to increased host galaxy contamination in the short wavelength filters if they were matched to the resolution of F770W. 

A system is considered a dual AGN candidate if the \texttt{SKIRTOR} AGN model outperforms both the stellar model and the stellar+dust model with $\rm \Delta BIC>10$. Out of 197 companions, 100 show a statistically significant improvement. However, we noticed that some improvements in BIC values occur because the companion is a foreground or background source, which prevents an accurate SED fit when the redshift is fixed to that of the central AGN. Adding an additional AGN template makes the SED shape more flexible and thus improves the fit. Therefore, the dual AGN candidates identified here are provisional, and their physical association will be further evaluated using the full probability distribution of \zphot\ measured in Section \ref{subsec:photoz}.  

It is also noteworthy that companions with an insignificant $\rm \Delta BIC$ could still harbor lower-luminosity or obscured AGNs. These AGNs might have SED signatures concealed by the host galaxy or the dusty torus, particularly in the case of high redshift, faint systems. In these scenarios, the longest NIRCam filter probes only the rest-frame optical emissions, which could be significantly attenuated.

\subsection{Photometric redshift and stellar mass estimates}
\label{subsec:photoz}
We construct multiband SEDs (corrected for galactic extinction) using deblended HST+JWST photometry derived from the decomposed \ss\ component for both the AGN host and its companions. Our objective is to measure the full probability distribution of \zphot\ and evaluate the physical association of projected pairs. An additional 0.1~mag uncertainty is added in quadrature to the magnitude uncertainty of the AGN host output by \texttt{GALFITM} to account for systematic uncertainties in AGN-host decomposition, as commonly implemented in the literature \citep[e.g.,][]{Zhuang2024}. For the non-AGN companions, we add an additional 0.02 mag uncertainty \citep[e.g.,][]{Weaver2022}. 

We employ \texttt{LePhare} for the \zphot\ measurements and use the same set of galaxy templates as in \cite{Ilbert2013}, as the potential AGN contamination has been subtracted using spatial decomposition. We adopt \texttt{Z\_ML} (i.e., the median of the redshift posterior) as our fiducial \zphot\, since it offers more reliable uncertainties and reduces the aliasing effect \citep{Ilbert2013}. With \zphot\ derived for each AGN and their companions, and \zspec\ available for a subset of AGNs, we further estimate their stellar mass using \texttt{CIGALE} based on the decomposed \ss\ flux and the stellar templates adopted in Section \ref{subsec:second}. 

Since most companions with subarcsec separations are either not effectively deblended from the central AGN in the COSMOS2020 catalog \citep{Weaver2022} or are too faint to be detected by ground-based facilities, the \zphot\ derived solely from the five broadband HST+JWST imaging is inevitably uncertain. To evaluate the accuracy of the derived \zphot, we compare them with reliable \zspec\ values for AGNs in our parent sample. Five objects with saturated NIRCam images, and thus unreliable decomposition results, are excluded from the comparison. We also restrict this exercise to AGNs with $f_{\rm gal}\gtrsim20\%$, as is the case for our pairs. The estimates of photo-$z$ using our method for point-source dominated systems would be more uncertain due to challenges in disentangling the host galaxy emissions. Their photo-$z$ can be more reliably measured from multiwavelength data, including ground-based photometry, using AGN-dominated templates \citep{Salvato2011}, where the contamination from nearby companions is negligible compared to the bright quasar emissions. Since such objects with $f_{\rm gal}<20\%$ account for only $\sim5\%$ of our final pair sample (Section \ref{sec:result}), their photo-$z$ uncertainty does not significantly impact our results. 

As shown in Figure \ref{fig:lephare}, even with photometry limited to five broadbands, we can still measure \zphot\ across a wide redshift range without bias on average. However, notable deviations occur for individual objects. Generally, we achieve an accuracy of $\sigma_{\Delta z}=0.07$ with an outlier fraction of $\eta=2.3\%$ at $\rm F814W\ (host\ only) < 22.5$~mag. For fainter objects, the accuracy decreases to $\sigma_{\Delta z}=0.11$, while the outlier fraction increases to $\eta=22.6\%$. 

\subsection{Pair Probability}
\label{subsec:pair}

\begin{figure}
\includegraphics[width=0.49\linewidth]{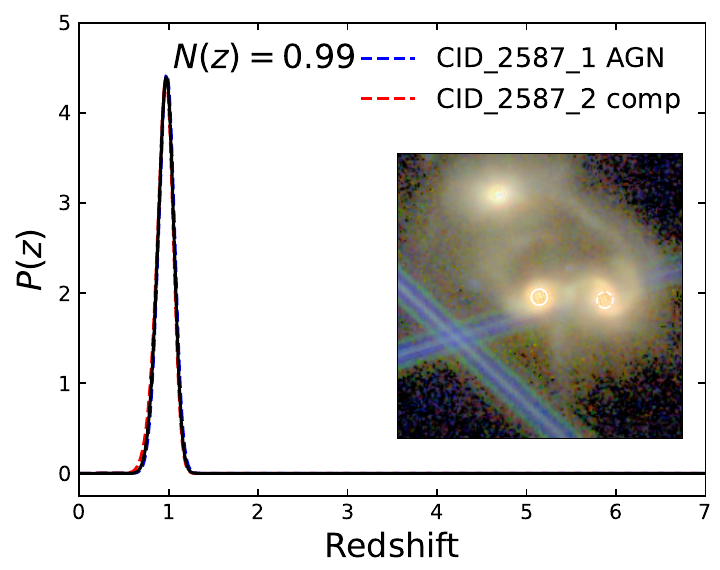}
\includegraphics[width=0.49\linewidth]{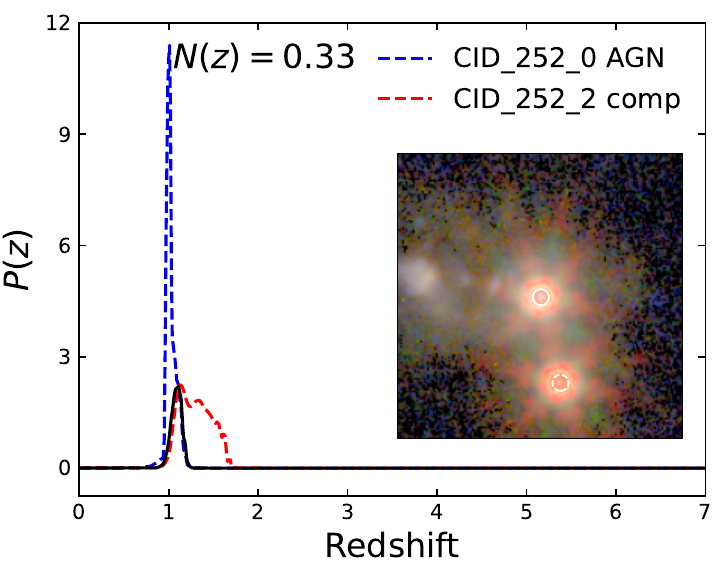}
\includegraphics[width=0.49\linewidth]{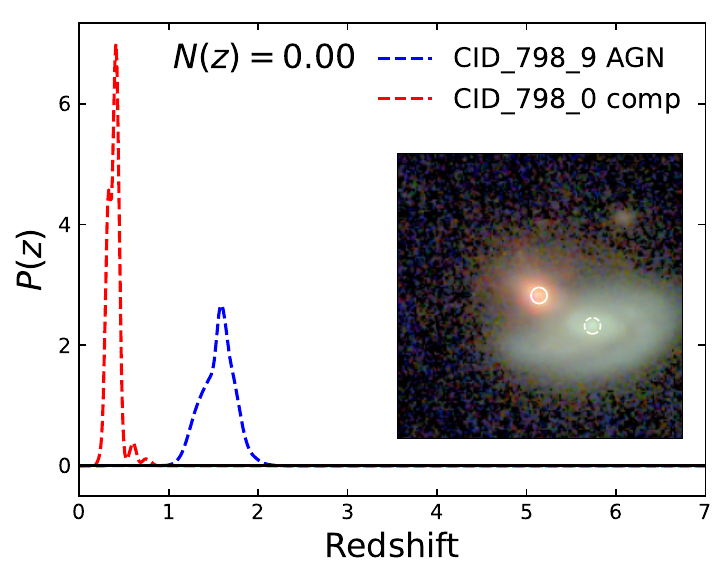}
\includegraphics[width=0.49\linewidth]{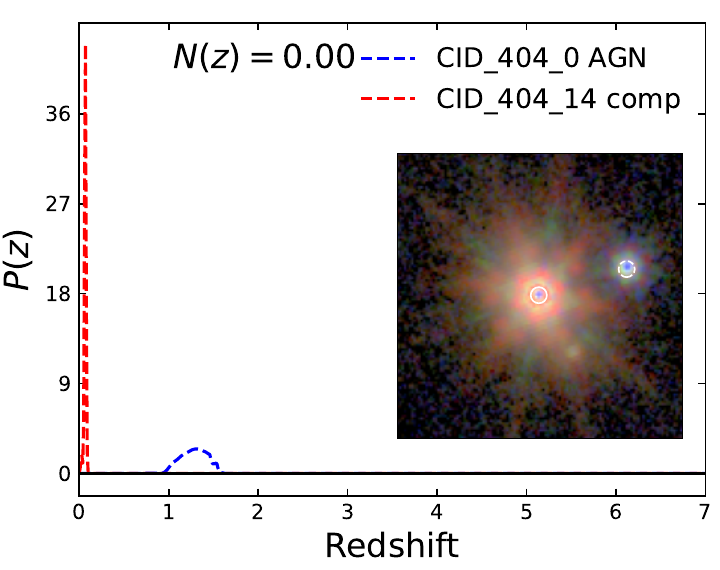}
\caption{Probability redshift distribution for four representative pairs. The blue and red dashed curves represent the PDZ for the central AGN (solid white circle) and its companion (dashed white circle), respectively. The solid black curve shows the combined probability of the pair, whose integral yields \nz. The inset shows the NIRCam color image of each pair.}
\label{fig:pdz}
\end{figure}

\begin{figure*}
\includegraphics[width=\linewidth]{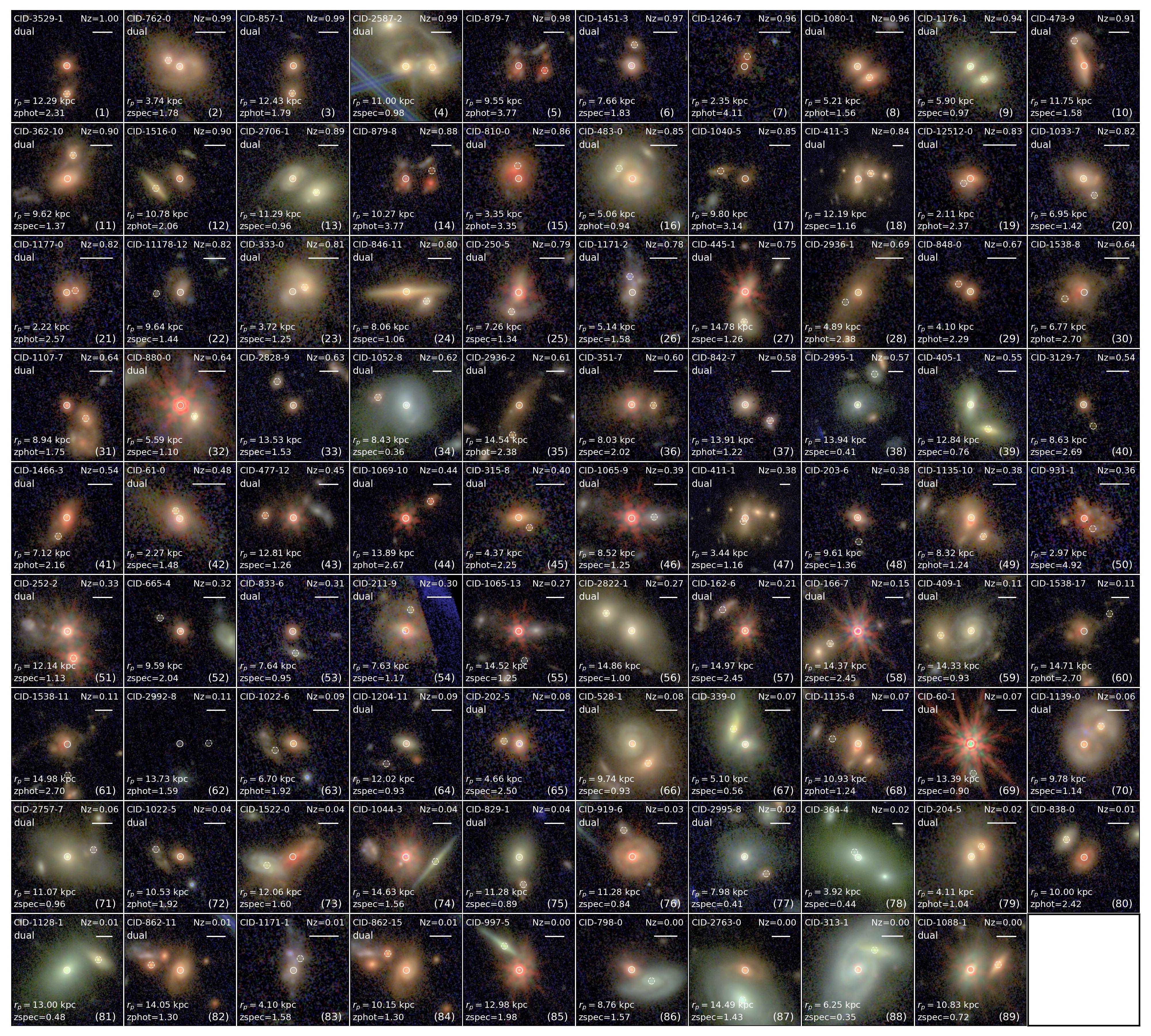}
\caption{Images of our final pair sample ($r_p < 15$ kpc, stellar mass ratio $>1:6$) in the order of declining \nz. North is up and east is left. Each panel displays the object name of the companion (white dashed circle) around the primary X-ray AGN (white solid circle in the center), the redshift of the primary AGN, the projected pair separation, and the probability that they share the same redshift (\nz). Systems identified as dual AGN candidates are labeled. The white horizontal line shows the $1\arcsec$ scale bar.} 
\label{fig:images}
\end{figure*}

\begin{figure*}[!t]
\includegraphics[width=\linewidth]{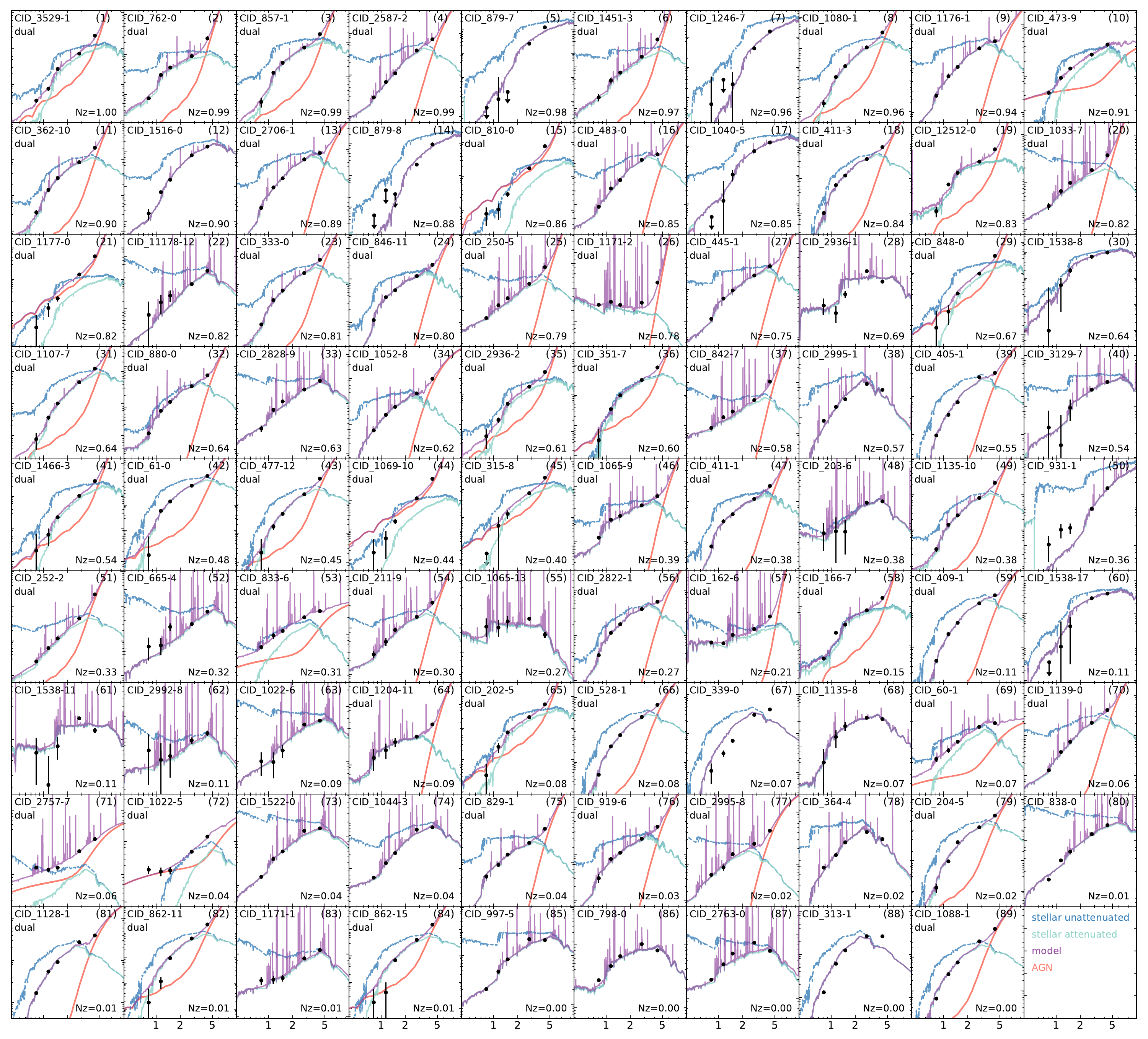}
\caption{A gallery of nuclear SEDs (plotted in logarithm scales on both axes) in units of flux density (mJy, arbitrary normalization) vs. observed wavelength (\um) for each companion in our final pair sample, in the same order as Figure \ref{fig:images}. The observed SEDs are shown as black circles. The best-fit SED template derived by fixing the redshift to that of the central AGN is shown in purple. Its components: unattenuated stellar emission (blue), attenuated stellar emission (green), and AGN emission (red), are displayed. Systems identified as dual AGN candidates (labeled below the object name) with high $N_z$ values are characterized by a significant emission excess in the F444W band, which cannot be explained by stellar and galactic-scale dust emissions, as evaluated through the $\Delta \rm BIC$ value. Systems with a low $N_z$ that also require an AGN template to fit the nuclear SED are likely projected interlopers whose SED shape cannot be fitted by stellar templates alone when the redshift is fixed to that of the central AGN. }
\label{fig:sed}
\end{figure*}

The combined redshift probability distribution function (PDZ) of the projected pair, defined as 
\begin{equation}
    P(z) = \frac{2\times P_a(z) \times P_c(z)}{P_a(z) + P_c(z)},
\end{equation}
can be measured from the PDZ of the AGN $p_a(z)$ and its companion $p_c(z)$. 
In this equation, $P_a(z)$ and $P_c(z)$ are broadened from the PDZ by convolving with a Gaussian kernel. The kernel's width is determined by the magnitude-dependent scatter of \zphot\ relative to \zspec\ (Figure \ref{fig:lephare}), subtracting the median \zphot\ error in each magnitude bin in quadrature. This procedure accounts for the statistical scatter of \zphot\ relative to \zspec\ caused by systematic effects such as template choices and the inaccurate assessment of photometric uncertainties in image decomposition. These factors are not captured by the PDZ of each individual source, which is solely determined from the fitting goodness based on the limited input galaxy templates. 
The normalization, $M(z) = (P_a(z)+P_c(z)) / 2$, is constructed such that $\int_{0}^{\infty} M(z) = 1$ \citep{Duncan2019}. The effective number of physically associated true pairs at redshift $z$ for a given projected pair can be derived as
\begin{equation}
    N_z = \int_{0}^{\infty} P(z) dz.
\label{eq:Nz}
\end{equation}

In the above calculations, we utilize the PDZ measured from our decomposed host galaxy photometry for the AGN, instead of relying on \zspec\ or the PDZ from \cite{Marchesi2016}. This choice ensures consistency in measuring the PDZ and its uncertainty for both the AGN and its companions, thereby minimizing potential systematic effects. Additionally, there is still some ambiguity as to which nucleus the measured \zspec\ belongs to in close pairs. Nevertheless, we verify that using $\zspec$ (broadened by the typical \zphot\ uncertainty) leads to consistent results in the statistical analysis of the pair sample (e.g., the pair fraction presented in Section \ref{sec:result} changes only by a factor of $\sim1.08$), although the estimated \nz\ for each individual system might differ.  

As some AGNs may have more than one projected companion within a radius of $\sim15$~kpc (i.e., 197 companions around 163 AGNs), we measure \nz\ for each projected pair separately and sum them to derive the total pair counts \citep[e.g.,][]{Duncan2019}. However, the summed \nz\ for each AGN system cannot exceed one (i.e., triple and multiple systems are treated as one pair). In Figure \ref{fig:pdz}, we present $P(z)$ for four representative pairs. The first pair, displaying spectacular tidal features, is also conclusively revealed to be a genuine merging pair using our probabilistic counting method with $N_z=0.99$. The second pair, CID-252, is identified as a dual AGN through imaging and nuclear SED analysis (Sections \ref{subsec:galfitm} and \ref{subsec:second}). Despite the presence of two bright point sources complicating the \zphot\ measurements, there is still a significant probability, $N_z=0.33$, that the two nuclei reside at the same redshift. In fact, the two nuclei have very similar photo-$z$ values ($\zphot=1.003$ for CID-252-0 and $\zphot=1.092$ for CID-252-2) that are close to the $\zspec=1.134$ available for CID-252-0. The third pair represents a superposition of systems located at different redshifts. The fourth pair, previously identified as a dual AGN candidate by \cite{Stemo2021}, is revealed to be a superposition of an AGN and a foreground star, with the companion's SED best described by a star template in \texttt{LePhare}. 

In fact, since \cite{Stemo2021} selected dual and offset AGN candidates based solely on the presence of companion ``bulges'' within $r_p\sim20$~kpc of the central AGN in single-band F814W imaging, their sample is susceptible to significant contamination from foreground and background interlopers. To assess the reliability of their catalog, we examined 22 of their X-ray dual and offset AGN candidates that are covered in the current COSMOS-Web footprint. Among these, 21 sources have spectroscopic redshifts available for the central AGN. Notably, 5 of them -- CID-371, CID-378, CID-402, CID-404, and CID-1098 -- were found to be AGN-star superpositions. The companion galaxies in another 5 systems -- CID-366, CID-377, CID-471, CID-997, and CID-1179 (4 of which have \zspec\ for both the AGN and the companion) -- are clearly at different redshifts.
The remaining objects display varying probabilities ($\nz \sim 0.05-0.99$) of being genuine pairs. Consequently, the contamination rate of this sample is at least $45\%$ (10/22). Therefore, the results in \cite{Stemo2021} regarding the statistical assessment of the dual and offset AGN population and their connection with mergers should be treated with great caution.

\section{Results}
\label{sec:result}

We have measured the physical properties of 163 X-ray AGNs and their 197 companions, which have a host galaxy magnitude difference of no more than 3 mag in the F277W band (flux ratio $>1:15$), to assess their physical association and potential as dual AGNs. Given the challenges in reliably measuring \zphot\ and stellar masses for faint companions \citep[e.g.,][]{Kauffmann2020}, as well as in identifying the presence of a secondary AGN, we have applied a final criterion that the stellar mass ratio between the AGN and its companion galaxy need to exceed $1:6$. Adopting a threshold for the merging mass ratio, rather than a flux-based criterion, also facilitates comparison with cosmological simulations (Section \ref{subsec:simulation}). Companions classified as stars by \texttt{LePhare} are excluded. 

As a result, our final sample includes 89 projected pairs around 78 AGNs, which are likely in the late stage ($r_p<15$ kpc) of major mergers. Based on the multiwavelength morphology of our pairs and the non-detection of a foreground lens galaxy in deep NIRCam images, we conclude that none of the pairs are lensed AGNs. The physical properties of each pair are presented in Table \ref{table:decomp}. The statistics of the final pair sample is summarized in Table~\ref{tab:number}. Most pairs are X-ray obscured by $\nh>10^{22}\ \cm$.

\begin{table*}
\caption{Information and properties of the 78 AGNs and their 89 companions in our final pair sample. }
\centering
\begin{tabular}{cccccccccccc}
\hline
\hline
ID & Class & \zspec & Z\_ML & RA & DEC & $\logm$ & $\loglbol$ & ${\rm log}\,N_{\rm H}$ & \nz\ & $r_p$ & isdual\\
& & & & $^\circ$ & $^\circ$ & $\msun$ & $\ergs$ & \cm\ & & kpc & \\
(1) & (2) & (3) & (4) & (5) & (6) & (7) & (8) & (9) & (10) & (11) & (12)\\
\hline
CID-1246-8 & AGN & -99.000 & 4.109 & 150.015921 & 2.441975 & 10.36 & 45.51 & 23.28 & -- & -- & --\\
CID-1246-7 & comp & -99.000 & 4.394 & 150.015896 & 2.442067 & 10.39 & -- & -- &  0.96 &  2.35 & False\\
CID-3529-0 & AGN & -99.000 & 2.314 & 149.830108 & 1.990348 & 10.84 & 45.05 & 23.09 & -- & -- & --\\
CID-3529-1 & comp & -99.000 & 2.350 & 149.830108 & 1.989932 & 10.65 & -- & -- &  1.00 & 12.29 & True\\
\hline
\end{tabular}
\tablecomments{Col. (1): Object name based on the combined Chandra ID of the X-ray AGN and the object ID in our \sex\ source detection catalog. The primary X-ray AGN is not necessarily labelled as source ``0''. Col. (2): Whether the source is the central X-ray AGN (``AGN'') or its companion (``comp''). Col. (3): Reliable spectroscopic redshifts from \cite{Marchesi2016}, if available. Col. (4): Photometric redshifts measured from \texttt{LePhare}. Cols. (5-6): Right ascension and declination in the F277W band. Col. (7): Stellar mass. Col. (8): Bolometric luminosity. Col. (9): Column density. Col. (10): Effective pair count for each companion. Col. (11): Projected separation. Col. (12): Whether the object is identified as a dual AGN candidate. This table is published in its entirety in the machine-readable format.}
\label{table:decomp}
\end{table*}

\begin{table}
    \centering
\renewcommand{\arraystretch}{1.2}

    \begin{tabular}{ccc}
    \hline
    \hline
        $\nh<10^{22}\ \cm$ & $\nh>10^{22}\ \cm$ & no $\nh$ \\
        \hline
        27 AGN & 41 AGN & 10 AGN\\
    \hline
     \noalign{\vskip 10pt} 

    \hline
    \hline
    \zspec & \zphot & total \\
    \hline
    46 AGN & 32 AGN  & 78 AGN\\
    \hline
    \noalign{\vskip 10pt} 

    \hline
    \hline
    dual & offset & total\\
    \hline
    59 pairs & 30 pairs & 89 pairs \\
    \hline

    \noalign{\vskip 10pt}    
    \hline
    \hline 
    genuine dual & genuine offset & genuine total\\ 
    \hline
    28 pairs & 10 pairs &  38 pairs\\
    \hline
    \end{tabular}
    \caption{Statistics of the final pair sample (89 companions around 78 AGNs), divided into (1). X-ray unobscured and obscured AGN, (2). redshift type for the X-ray AGN, (3) dual and offset AGN candidates, and (4) ``genuine'' dual and offset AGN candidates derived by summing $N_z$ for each projected pair.}
    \label{tab:number}
\end{table}

The images of these pairs, sorted by \nz, are presented in Figure \ref{fig:images}. The nucleus-region SEDs of the 89 companions are displayed in Figure \ref{fig:sed}. Among them, 59 pairs are potential dual AGN candidates, based on the requirement of a second AGN to explain the companion's nuclear SED when fixing the redshift to that of the central AGN. Most of the dual AGN candidates are best fit by a type 2 AGN template ($\theta = 70^\circ$) with negligible contribution to short wavelength filters. However, this is mostly due to our adoption of very flexible stellar templates that can mimic a variety of observed SEDs, but only two AGN templates with varying normalizations in the fitting. This approach is sufficient to reveal any excess emission in the F444W band that cannot be explained by stellar and galactic-scale dust emissions, but it cannot constrain the AGN SED across the entire wavelength range, especially when the AGN is not dominant.

\begin{figure}
\includegraphics[width=\linewidth]{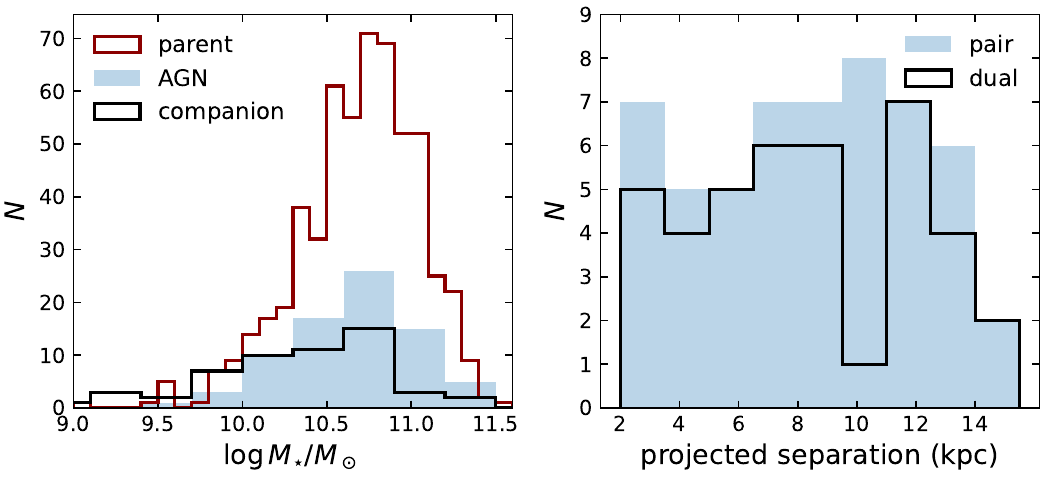}
\caption{Left: distribution of stellar mass for the central AGN (blue) and its companions (black) for the final pair sample. The stellar mass distribution for the parent sample of 571 AGNs is shown in red. Right: distribution of projected separation for pairs and dual AGN candidates with $N_z>0.3$. }
\label{fig:mass}
\end{figure}

\subsection{Statistical properties of the pair sample}
The total number of \textit{genuine} physical pairs (including both dual and offset AGNs) in our sample, measured by summing \nz\ for the 89 projected pairs, is $\sim38$, corresponding to a pair fraction of $\sim6.6\%$ (38/571). The predicted number of \textit{genuine} dual AGNs is $\sim28$ when considering the full PDZ of the 59 candidates, and the number of \textit{genuine} offset AGNs is $\sim10$. The dual-to-offset ratio and their dependence on the host galaxy properties provide important clues about the triggering conditions of AGNs in mergers \citep[e.g.,][]{Capelo2017}. The high ratio ($\sim2.8$) observed in our sample, if confirmed, would imply that SMBHs in close major-merger pairs are preferentially triggered simultaneously, as reported in \cite{Perna2023} at $z\sim3$ and also seen in the local universe \citep{Koss2012}. Spatially resolved spectroscopic observations of AGN diagnostic lines, preferably in the rest-frame NIR wavelengths that are less affected by dust extinction \cite[e.g.,][]{Calabro2023}, are imperative to firmly distinguish between the dual and offset scenarios for our sample. 

The distribution of stellar mass and projected separation for the 54 pairs with $N_z>0.3$ that are likely physically associated is shown in Figure \ref{fig:mass}. The stellar mass of AGN in pairs appears similar to that of the parent X-ray AGN population. Overall, our sample probes relatively massive dual and offset AGNs in the poorly explored kpc-scale regime, extending down to $r_p \sim 2$~kpc. 

Interestingly, we do not observe an increase in the number of pairs ($N_{\rm pair}$) and dual AGNs as the separation decreases (limited to pairs with $N_z>0.3$). The \lbol\ of the central AGN shows a tentative increase at small \rp, but the enhancement is statistically insignificant, with a Spearman's correlation coefficient of $\rho=-0.19$ and a $p$-value of 0.19. This contrasts with the significant increasing trend reported in several local studies, albeit over a much broader separation scale (e.g., \citealt{Liu2011a, Liu2011, Ellison2011, Koss2012}; but see \citealt{Hou2020}). Since the occurrence of AGN pairs as a function of separation depends on both the AGN duty cycle in mergers and the dynamical friction timescale, changes in either across cosmic times, arising from intrinsic evolution or the varying luminosity range probed by high redshift and local samples, could cause the distinct trends observed in the high-redshift and local universe. 

One plausible explanation for the lack of a separation-dependent trend is the ample cold gas reservoir at high redshifts. While mergers might enhance gas infall likelihood, the accreting luminosity may not be significantly boosted compared to single AGNs triggered by internal instabilities. Moreover, feedback from high-redshift, luminous AGNs might be more efficient in self-regulating their luminosity. However, potential issues, such as the small sample size, use of projected separations, non-negligible contamination from unassociated pairs (especially at large separations), substantial measurement uncertainties and variability in \lx, could erase potential correlations that are generally weak in simulations \citep{Chen2023}. 

On the other hand, none of the pairs have a separation below $\sim2$ kpc, as confirmed by our visual inspection of PSF-subtracted images for the parent AGN sample. This finding is puzzling, given that the resolution in F277W (PSF FWHM $\sim0.\arcsec11$) should allow for resolving pairs down to $\sim1.0$ kpc, even up to the highest redshift probed by our sample. One possible explanation is that pairs at such close separations may be significantly more obscured, rendering them undetectable in F277W while being unresolvable in F444W (PSF FWHM $\sim0.\arcsec16$).  However, we do not observe a significant correlation between \nh\ and \rp\ for systems with $N_z>0.3$ (Spearman's $\rho=-0.13$ and $p$-value = 0.41), although we caution that many systems only have a few counts in the $0.5-8$~keV band, which are insufficient to reliably constrain \nh\ and the absorption corrected \lx. Besides, the most heavily obscured Compton-thick AGNs are largely missing in the Chandra surveys \citep[e.g.,][]{Li2019}. Moreover, the global \nh\ in high-redshift galaxies is much higher than in the local universe \citep[e.g.,][]{Circosta2019, Gilli2022}, such that the enhancement in \nh\ induced by mergers may not be significant. 

Alternatively, the lack of pairs at $r_p<2$~kpc may suggest that galaxies decay and coalesce more rapidly at small separations than predicted by the classic Chandrasekhar dynamical friction model, where the dynamical friction timescale is proportional to the BH separation ($t_{\rm dyn}/{\rm yr} \sim 10^4 \Delta r/{\rm pc}$ for a $\sim10^8\ \msun$ BH; e.g., \citealt[e.g.,][]{Yu2002, Chen2020dyn}). This would result in a roughly constant AGN pair fraction in linear separation bins if the AGN duty cycle remains unchanged during mergers. 
It is worth noting that the fuzzy dark matter model predicts a pileup of pairs at $r_p<2$~kpc due to the core stalling effect, resulting in a much longer dynamical friction timescale than the classic Chandrasekhar formula \citep[e.g.,][]{Hui2017}. However, this effect is not observed in our sample.  It is also possible that the secondary galaxy undergoes significant tidal disruption in the final coalescence stage, making it not readily observable as a separate core and also suppressing the triggering of its central SMBH.

\subsection{Observed pair fraction and its evolution}
\label{subsec:fraction}

\begin{figure*}
\includegraphics[width=0.49\linewidth]{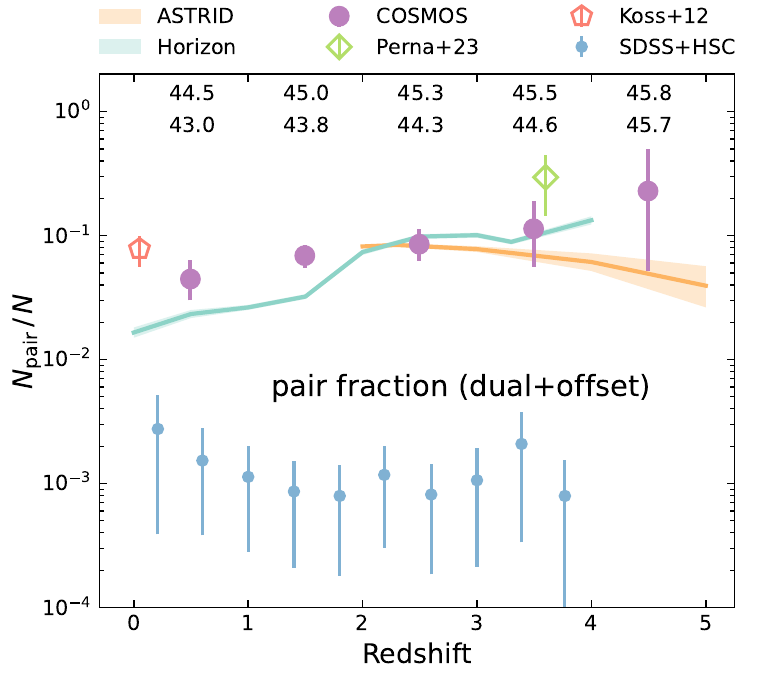}
\includegraphics[width=0.49\linewidth]{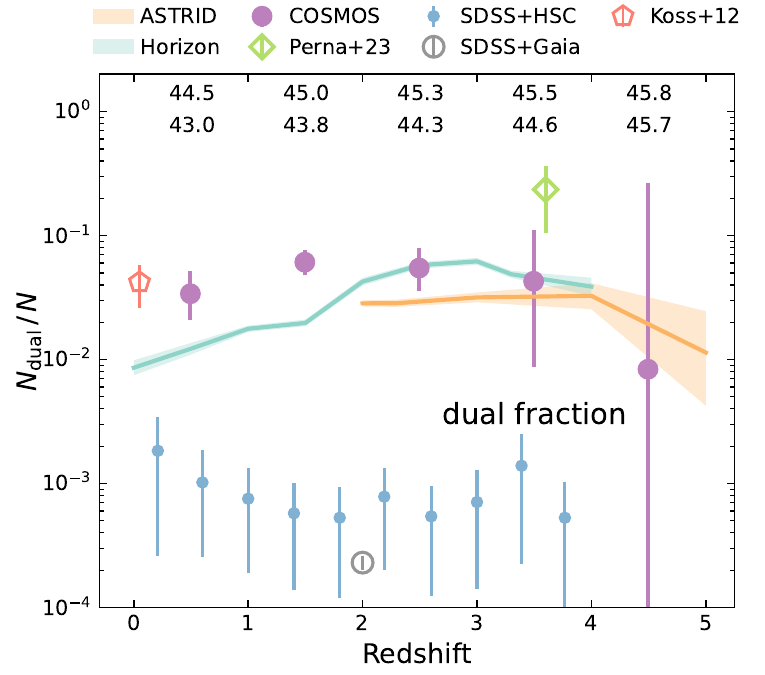}
\caption{Dual and offset AGN fraction with $r_p<15$~kpc and a merger mass ratio $>1:6$ as a function of redshift for our final pair sample. The left panel shows the total pair fraction (i.e., dual + offset), while the right panel displays the dual AGN fraction (purple circles). The luminosity limit and median luminosity in each redshift bin is labeled. For comparison, we include the dual and pair fraction from the \texttt{ASTRID} and \texttt{Horizon-AGN} cosmological simulations after matching with our sample selection criteria. Additionally, pair and dual AGN fraction based on local ultra hard X-ray selected AGNs \citep{Koss2012}, as well as unobscured quasars in SDSS+HSC (\citealt{Silverman2020}; corrected for the spectroscopic confirmation rate and incompleteness at small separations) and SDSS+Gaia (\citealt{Shen2023}; corrected for the lensed quasar fraction) with $r_p < 15$ kpc but different luminosity thresholds (see Section \ref{subsec:fraction} for details) are plotted to illustrate the pronounced dependence of the pair fraction on AGN luminosity and obscuration.}
\label{fig:fraction}
\end{figure*}

Given $P(z)$ for each projected pair and $P_a(z)$ for each AGN in our parent sample (i.e., the 571 X-ray AGN), we can calculate the fraction of \textit{genuine} dual and offset AGNs relative to the overall AGN population as a function of redshift by integrating and computing their ratios. The redshift dependence of the pair and dual AGN fraction, along with the luminosity limit of the central AGN in each redshift bin, is presented in Figure \ref{fig:fraction}. The error bars are computed using the \cite{Gehrels1986} approximation. 

We find an overall increase in the pair fraction (\fp) with redshift, rising from $\fp\sim4.5_{-1.5}^{+1.9}\%$ at $z\sim0.5$ to $\fp\sim22.9_{-17.7}^{+27.5}\%$ at $z\sim4.5$. This may signify the increased galaxy merger rate at high redshift \citep[e.g.,][]{Duncan2019}. In contrast, the dual AGN fraction (\fd) shows a tentative decrease with redshift, exhibiting a significant drop in dual AGN candidates at $z>4$. Since the dual AGN fraction (relative to the parent AGN population) is proportional to the duty cycle of the secondary AGN, the declining trend may suggest a reduced duty cycle of high-redshift luminous (due to the detection limit) AGNs compared to low-redshift, lower-luminosity ones \citep[e.g.,][]{DeGraf2017, Trebitsch2019}. Additionally, the small number of objects in the parent sample and the challenge of identifying a secondary AGN at high redshifts could contribute to the apparent deficiency of dual AGNs at $z>3$, as the F444W band shifts to rest-frame wavelengths below $\rm \sim1\ \um$. 

Figure \ref{fig:fraction} compares our pair and dual AGN fractions with literature results across a broad redshift range. In the local universe, our results generally align with those of \cite{Koss2012}, who reported a pair fraction of $\sim9.6\%$ and a dual fraction of $\sim4.8\%$ at $r_p<15$~kpc in a sample of ultra hard X-ray selected AGNs. Remarkably, at higher redshifts, our \fd\ exceeds that of luminous and unobscured dual quasar candidates selected from SDSS+Gaia {($\loglbol > 45.8\ \ergs$ for both nuclei, which are $\gtrsim1.5$ dex shallower than our sample)} by two orders of magnitude at $z\sim2$ and $r_p<1.\arcsec8$ ($\sim15$~kpc; \citealt{Shen2023})\footnote{The SDSS double quasars in \citealt{Shen2023} could be dual or lensed quasars, and we assume dual : lens = 1 : 1 to derive the dual fraction.}. 
Similarly, the \fp\ and \fd\ derived from pairs selected by having an HSC companion around an SDSS quasar {($\loglbol>45.5\ \ergs$ at $z\sim2$ for the central quasar and flux ratio $>1:10$)} within $r_p < 15$~kpc are $\sim1-2$ orders of magnitude lower than our measurements \citep{Silverman2020}\footnote{The \fd\ in \cite{Silverman2020} is derived by multiplying the observed fraction of projected pairs with a success rate of 50\% in spectroscopic follow-up observations (3 dual quasars out of 6 observed sources) and corrected for the incompleteness at $0.\arcsec6 < r_p<2.\arcsec5$ by assuming a flat distribution in separation. Here, we update the success rate based on their follow-up study in \cite{Tang2021}, which identifies 3 dual quasars and 3 offset quasars out of 26 additional sources, to derive \fp\ and \fd. Additionally, their sample only probes $r_p\gtrsim5$~kpc ($\sim0.\arcsec6$). We assume that their flat separation distribution extends to $r_p<5$~kpc to derive the pair fraction at $r_p<15$~kpc.}. 

On the other hand, \cite{Perna2023} conducted NIRSpec IFU observations of 17 AGNs at $2<z<6$. These AGNs span $\lbol \sim 10^{44}-10^{46}\ \ergs$, a range similar to our sample. They reported five physically associated pairs at $z\sim3.5$, including three dual AGNs, a triple AGN, and a dual candidate, with $3<r_p<28$~kpc. Out of the four Chandra detected pairs, three are X-ray obscured, and the remaining X-ray undetected pair is optically classified as type 2.  This high confirmation rate corresponds to $\fp\sim29\%$ and $\fd\sim23\%$, which roughly align with our pair fraction at $z>3$ (Figure \ref{fig:fraction}). 

The apparent discrepancy in the pair and dual AGN fraction between moderate-luminosity X-ray AGNs and luminous SDSS type 1 quasars can be attributed to their distinct luminosity and obscuration levels (Figure \ref{fig:lumin}). Theoretical models predict that the duty cycle of obscured, moderate-luminosity accretion can be up to $\sim100$ times larger than that of the unobscured, luminous quasar phase \citep[e.g.,][]{Blecha2018, Trebitsch2019}.  This decreasing duty cycle with luminosity would notably reduce the dual fraction among luminous quasars if the quasar duty cycle is not significantly boosted in mergers \citep[e.g.,][]{Shen2023}. Additionally, if gas-rich mergers at high redshift preferentially produce obscured AGNs, these pairs would elude the optical SDSS selection, leading to much suppressed fractions of dual and offset quasars among the SDSS sample as well. 

Another intriguing finding is that both observations and simulations shown in Figure \ref{fig:fraction} reveal a comparable dual and offset fraction among AGNs. This suggests that the probability of the companion being an AGN is significantly enhanced compared to the overall AGN duty cycle. For example, if the typical AGN duty cycle is $\sim0.2-0.3$ \citep[e.g.,][]{Trebitsch2019, DeGraf2017} within the luminosity range of our companions (assumed to be one dex below the primary AGN; see Section \ref{subsec:simulation}), and both members in mergers are activated stochastically, we would expect to find $\sim2-4$ times more offset AGNs than dual AGNs. However, we observe $\sim2.8$ times more dual AGNs than offset AGNs.

However, we acknowledge that both this study and \cite{Perna2023} are still subject to small number statistics and potential biases. The sample in \cite{Perna2023} appears to include a high number of X-ray obscured AGNs. Nevertheless, there is still an abundant population of type~1 AGNs at $z\sim3$ that apparently lacking any companions in deep NIRCam images \citep{Zhuang2024}, as well as X-ray unobscured pairs (Figure \ref{fig:lumin}). Our identification of dual AGNs at high redshift is likely incomplete, and the ``genuine'' pair sample may still suffer from contamination from projected pairs or pairs that are not yet interacting. X-ray-selected AGNs are also biased against the most heavily obscured Compton-thick AGNs \citep[e.g.,][]{Lanzuisi2018, Li2019}, which might be prevalent in mergers \citep[e.g.,][]{Kocevski2015, Ricci2017, Li2020, Ricci2021}. The impact of a potentially missing Compton-thick population on understanding the AGN duty cycle in mergers will be addressed in our future work. Further spectroscopic observations of a larger sample, covering broad parameter spaces in \lbol\ and \nh\ (especially in uncovering heavily obscured AGNs with IR selection techniques), are imperative to confirm the increased pair fraction in the early universe and to reveal their triggering conditions.

\subsection{Comparison with cosmological simulations}
\label{subsec:simulation}

In theoretical models, the occurrence of dual and offset AGNs depends on various factors, including the merging mass ratio, the structure and kinematics of the host galaxy, the gas content, distribution, and accretion history, the adopted luminosity threshold, and the detailed sub-grid physics regarding BH fueling and feedback mechanisms \citep[e.g.,][]{Steinborn2016, Capelo2017, LiK2021, Volonteri2022}. It is thus of great interest to compare the observational constraints on the incidence of AGN pairs with simulation results across cosmic times to examine model details \citep[e.g.,][]{Silverman2020, Shen2023}. 

As extensively discussed by \cite{DeRosa2019}, comparing with simulations requires careful matching of sample properties. To minimize discrepancies, we match our sample selection with the \texttt{ASTRID} and the \texttt{Horizon-AGN} simulations. \texttt{ASTRID} is a large volume (250 Mpc/$h$) cosmological simulation with a spatial resolution of 1.5 ckpc/$h$ and a current final redshift of $z=2$ \citep{Bird2022, Ni2022}. It employs state-of-the-art dynamical-friction modeling to track SMBH pairs down to a proper separation of $\Delta r \lesssim 1$ kpc, a regime previously achievable only in idealized merger simulations \citep{Chen2022dyn, Chen2023}. \texttt{Horizon-AGN} has a smaller simulation volume of 100~Mpc/$h$ with a spatial resolution of 1~kpc \citep{Dubois2014}. It was the first large-scale cosmological simulation to include a gas dynamical friction model, rather than artificially repositioning SMBHs, avoiding spurious kicks and oscillations of SMBH due to external perturbations and finite resolution effects. In \texttt{Horizon-AGN}, SMBHs merge at a proper separation $\Delta r < 4$~kpc, resulting in a lack of pairs at separations below this threshold and an underestimation of the pair fraction therein. The merging criterion in \texttt{ASTRID} is $\Delta r<3~{\rm kpc}/h/(1+z)$, capturing more small-separation pairs at high redshift, while at low redshift, the two simulations become comparable in terms of the distance to which SMBHs can be tracked.

As shown in Figure \ref{fig:lumin}b,  our AGN sample occupies a similar region on the \lbol--\nh\ plane as those in \texttt{ASTRID} after matching in the detection limit in \lbol\ as a function of redshift. Note that in \texttt{ASTRID}, \nh\ reflects only galactic-scale obscuration, measured as the median value across 48 lines-of-sight for each AGN. In contrast, both torus and galaxy obscuration contribute to the line-of-sight \nh\ of our X-ray AGNs, with the latter likely playing a significant role (reaching $\nh > 10^{23}\ \cm$) at high redshift \citep{Circosta2019, Gilli2022}. Most other high redshift dual AGN searches target only the most luminous and unobscured quasar pairs, a population not representative in mergers and not well sampled in simulations \citep{ChenY2023, Shen2023}. Therefore, our sample provides the first opportunity for a direct comparison of the incidence of kpc-scale dual and offset AGNs across a wide range of redshifts between observations and theoretical models.

We apply the same criteria: projected separation $r_p<15$~kpc, galaxy merger mass ratio $>1:6$, and the detection limit in \lbol\ of the central AGN in Figure~\ref{fig:lumin}a to select simulated pairs. Multiple BHs in simulations within $r_p<15$~kpc are counted only once. 
The simulated dual AGNs are selected by setting a luminosity limit on the secondary AGN to be one dex below that of the central AGN. This luminosity ratio is similar to that observed in \cite{Perna2023} and is adopted based on the argument that the secondary AGN should still be relatively luminous to be identifiable in the SED, yet less luminous than the central AGN given their flux ratio.

Before proceeding to detailed comparisons, we note that the definitions of dual and offset AGNs in simulations and observations differ in some aspects. In observations, dual AGNs identified in JWST imaging are associated with their own ``bulge'' (i.e., a centrally concentrated \ss\ model with the same center as the PSF model) by selection (i.e., we require the stellar mass ratio $>1:6$). For offset AGNs, the ``bulge'' of the companion galaxy is clearly visible, which we used to infer the presence of an inactive BH assuming a 100\% BH occupation fraction. Additionally, none of the companions in our parent pair sample (i.e., 197 pairs with a flux ratio $>1:15$) appear to be  compact point sources lacking its own host-galaxy component based on visual inspection. Consequently, we did not observe wandering BHs \citep[e.g.,][]{Volonteri2016, Tremmel2018} in our sample.  However, this study is not dedicated to searching for such systems, which we will address in future work. 

In \texttt{ASTRID} and \texttt{Horizon-AGN}, BH pairs are first identified and then assigned to host galaxies using \texttt{Subfind} \citep{Springel2001} or \texttt{HaloMaker} \citep{Aubert2004}. At large separations, each BH may be assigned to different galaxies. However, during close encounters,  the halo finder algorithms may not isolate merging systems and may attribute both BHs to the same host galaxy, even if two ``bulges'' are discernable from imaging (Case I). On the other hand, BH pairs may be found in the same galaxy after the two galaxies have merged (Case II). The second ``bulge'' may be heavily tidally stripped, making it indistinguishable from the disturbed disk of the primary galaxy. Consequently,  secondary BHs can appear as wandering and naked, deposited on outer orbits of the primary galaxy due to inefficient dynamical fraction \citep[e.g.,][]{Volonteri2016, Tremmel2018, Chen2023}

When counting dual and offset AGNs in \texttt{Horizon-AGN}, we included cases where BH pairs are hosted by the same galaxy to avoid missing the ``two-bulge'' cases not identified by \texttt{HaloMaker}. However, this also includes many wandering BHs \citep{Volonteri2022}. This mainly affects the comparison of the offset AGN fraction with observations, since the inactive secondary BHs cannot be identified by the presence of a ``bulge'' in JWST imaging. The inefficient accretion onto wandering secondary BHs makes them unlikely to shine as dual AGNs above our luminosity limit \citep[e.g.,][]{Volonteri2016}, so they do not  significantly affect the dual AGN fraction in simulations. 

We employ a correction factor to reduce contamination from wandering BHs. In \texttt{Horizon-AGN}, the ratio of Case I to Case II is $\sim50\%$ given the luminosity limit of the central AGN, so we multiply the total number of pairs by 0.5 for the ``same galaxy'' case to count the number of bona-fide pairs with the same definition of observations. Similarly, in \texttt{ASTRID}, Case II is overall negligible in the dual AGN catalog given our sample selection criteria, and we only include cases where the two BHs are in different subhalos to identify offset AGNs.

As shown in Figure \ref{fig:fraction}, the pair and dual AGN fractions in \texttt{ASTRID} generally align with our constraints at cosmic noon ($z\sim2-3$). However, there is a tentative deviation at the highest redshift bin ($z\sim4-5$), where \texttt{ASTRID} predicts a decline in the pair fraction, contrary to the observed increasing trend. Nevertheless, they remain consistent within the $1\ \sigma$ uncertainty region, given the substantial uncertainties in the observed pair fraction.  On the other hand, \texttt{Horizon-AGN} reproduces the observed redshift dependency for both \fp\ and \fd\ at $z<4$, although the pair and dual fractions are $\sim2$ times lower than observations at $z<1.5$. At $4<z<6$, none of the simulated AGNs in \texttt{Horizon-AGN} surpass our luminosity limit. 

In general, the tension between the observed incidence of unobscured quasar pairs (less than a few per thousand) and the predictions from simulations (a few percent) has been greatly alleviated by delving deeper into more typical, lower-luminosity AGNs, including obscured ones. The pair fraction from simulations and observations now agree within a factor of $\sim$2. This suggests that the overall prescriptions for galaxy merger rates, dynamics of SMBH pairs, and AGN triggering and feedback in simulations are reasonably accurate. However, certain discrepancies remain, and the uncertain luminosity limit for the observed secondary AGNs could also affect the dual AGN fraction in simulations by a factor of a few. Future spatially-resolved spectroscopic observations will allow  detailed examination of the stellar and gas kinematics in AGN pairs and their evolution in different merger stages, which will be instrumental to refine the subgrid physics in cosmological simulations.

\section{Concluding Remarks}
\label{sec:conclusion}

In this study, we use multiband HST and JWST NIRCam imaging data to systematically search for kpc-scale dual and offset AGNs across $z\sim0-5$ in late-stage ($r_p<15$ kpc) major galaxy majors (stellar mass ratio $>1:6$) within a parent sample of 571 X-ray AGNs. Multiwavelength simultaneous image decomposition and spatially-resolved SED analysis in the nuclear region are employed to deblend the photometry of X-ray AGNs and their surrounding companions, assess their physical associations, and identify potential secondary AGNs. 

Among the 89 massive companions detected around 78 X-ray AGNs, we estimate that $\sim38$ pairs are physically associated after accounting for the full redshift probabilistic distribution of each pair to minimize projection effects. Of these, $\sim28$ are predicted to be genuine dual AGNs, resulting in a dual-to-offset AGN ratio of $\sim2.8$. This higher occurrence rate of dual over offset AGNs implies a high probability of both BHs being active simultaneously in late-stage merging galaxy pairs. 

These promising dual and offset AGN candidates, whose luminosities and obscuration properties align with the dominant AGN population predicted by cosmological simulations, enable the first direct comparison with theoretical predictions on the incidence of AGNs in kpc-scale galaxy pairs up to $z\sim5$. The total dual and offset AGN fraction, coupled with specific sample selection criteria, tentatively rises toward higher redshifts, from $\sim4.5_{-1.5}^{+1.9}\%$ at $z\sim0.5$ to $22.9_{-17.7}^{+27.5}\%$ at $z\sim4.5$. The dual AGN fraction exhibits a tentative decrease at high redshifts, possibly reflecting the reduced duty cycle of high-redshift, luminous AGNs compared to their low-redshift, lower-luminosity counterparts. Additionally, the incompleteness in identifying secondary AGNs with imaging alone could contribute to the apparent deficiency of dual AGNs at $z>3$. Intriguingly, our observed pair and dual AGN fractions generally align with predictions from the \texttt{ASTRID} and \texttt{Horizon-AGN} cosmological simulations when matching sample selection criteria, and significantly exceed those measured from unobscured SDSS quasar pairs by $\sim2$ orders of magnitude. 

This work marks significant progress in the search for more typical (i.e., lower luminosity and obscured) dual and offset AGNs at high redshift and small separations, a regime that has just started to be explored systematically. Our analysis demonstrates that JWST's unparalleled sensitivity, spatial resolution, and wavelength coverage can deliver large samples of high-fidelity AGN pairs and provide meaningful constraints on their statistical properties and cosmic evolution, even with imaging data alone.  However, this approach cannot completely eliminate unassociated pairs. Spatially-resolved spectroscopy is crucial for measuring the redshift of our candidates and distinguishing between dual and offset scenarios \citep{Perna2023}. 

In general, pairs with a high \nz\ (Eq.~\ref{eq:Nz}), especially those showing evident merger signatures, are of the highest priority for spectroscopic follow-up. However, considering the outlier fraction of \zphot\ estimates for faint galaxies, pairs with a low \nz\ could still be genuine, while those with a high \nz\ might be interlopers. Ideally, spectroscopy of the entire sample (except for evident projected pairs) is desired to evaluate the  success rate of our imaging-based selection methods. This will also inform future searches for dual and offset AGNs in wide-area, multiband, high-resolution imaging surveys with additional capabilities of slitless spectroscopy, such as the Nancy Grace Roman Space Telescope and the China Space Station Telescope.

The fact that our sample probes obscured and/or lower-luminosity AGN pairs, the dominant population predicted by cosmological simulations, makes it ideal and much easier for revealing the stellar properties of the underlying host galaxies compared to unobscured dual quasar pairs. With the resolving power of JWST and ALMA, we can now systematically study host galaxy properties (e.g., structure, gas reservoir, stellar and gas kinematics, star formation rate and history) around each nucleus down to sub-kpc scales \citep{Chen2024, Ishikawa2024} for a statistically significant sample of high redshift dual and offset AGNs. 
Understanding their connection to SMBH activation in close pairs will provide crucial insights into the triggering conditions of AGN activity and allow us to examine predictions from merger models \citep{Steinborn2016, Chen2023}.

\begin{acknowledgments}
We thank Stefano Marchesi for providing us the X-ray spectral fitting catalog, and Mara Salvato for providing the probability distribution of photometric redshifts in \cite{Marchesi2016} for comparison. We thank Yelti Rosas-Guevara for useful discussions about dual AGNs in the EAGLE simulation. 
Based on observations with the NASA/ESA/CSA James Webb Space Telescope obtained from the Barbara A. Mikulski Archive at the Space Telescope Science Institute, which is operated by the Association of Universities for Research in Astronomy, Incorporated, under NASA contract NAS5-03127. Support for Program numbers JWST-GO-02057 and JWST-AR-03038 was provided through a grant from the STScI under NASA contract NAS5-03127. 
\end{acknowledgments}

\bibliography{sample631.bbl}
\bibliographystyle{aasjournal}

\end{document}